\documentclass[useAMS,usenatbib,usedcolumn]{mn2e}
\input psfig.sty
\usepackage{epsfig}
\usepackage{amsmath,amssymb}
\usepackage{psfig}
\usepackage{graphicx}
\usepackage{natbib}
\usepackage{ulem}

\newcommand{\kpc} {{\,\rm kpc}}

\newcommand{\kms}{{\,\rm {km\,s^{-1}} }}

\newcommand{\mpch}{\>h^{-1}{\rm {Mpc}}}   

\def \apj  {ApJ}

\def \apjl  {ApJL}
\def \prd {Phy.Rev.D}
\def \mnras {MNRAS}
\def \etal {et~al.~}\def \chisq  {\ifmmode  \chi^2   \else  $\chi^2$  \fi}  
\def \spose#1{\hbox  to 0pt{#1\hss}}  
\def \lta{\mathrel{\spose{\lower 3pt\hbox{$\sim$}}\raise  2.0pt\hbox{$<$}}}
\def \gta{\mathrel{\spose{\lower  3pt\hbox{$\sim$}}\raise 2.0pt\hbox{$>$}}}

\def \kms {\ifmmode  \,\rm km\,s^{-1} \else $\,\rm km\,s^{-1}  $ \fi }
\def \kpc {\ifmmode  {\rm kpc}  \else ${\rm  kpc}$ \fi  }  
\def \Msun {\ifmmode M_{\odot} \else $M_{\odot}$ \fi} 
\def \hMsun {\ifmmode h^{-1}\,\rm M_{\odot} \else $h^{-1}\,\rm M_{\odot}$ \fi}

\def \LCDM {\ifmmode \Lambda{\rm CDM} \else $\Lambda{\rm CDM}$ \fi}
\def \sig8 {\ifmmode \sigma_8 \else $\sigma_8$ \fi} 
\def \OmegaM {\ifmmode \Omega_{\rm M} \else $\Omega_{\rm M}$ \fi} 
\def \OmegaL {\ifmmode \Omega_{\rm \Lambda} \else $\Omega_{\rm \Lambda}$\fi} 
\def \Deltavir {\ifmmode \Delta_{\rm vir} \else $\Delta_{\rm vir}$ \fi}
\def \rhocrit {\ifmmode \rho_{\rm crit} \else $\rho_{\rm crit}$ \fi}
\def \rhou {\ifmmode \rho_{\rm u} \else $\rho_{\rm u}$ \fi}
\def \zc {\ifmmode z_{\rm c} \else $z_{\rm c}$ \fi}

\def \rhos {\ifmmode \rho_{\rm s} \else $\rho_{\rm s}$ \fi} 
\def \rs {\ifmmode r_{\rm s} \else $r_{\rm s}$ \fi} 
\def \cvir {\ifmmode c_{\rm vir} \else $c_{\rm vir}$ \fi} 
\def \Rvir {\ifmmode r_{\rm vir} \else $R_{\rm vir}$ \fi}
\def \Vvir {\ifmmode V_{\rm  vir} \else  $V_{\rm vir}$  \fi} 
\def \Mvir {\ifmmode M_{\rm  vir} \else $M_{\rm  vir}$ \fi}  
\def \Nvir {\ifmmode N_{\rm  vir} \else $N_{\rm  vir}$ \fi}  
\def \Jvir {\ifmmode J_{\rm vir} \else $J_{\rm vir}$ \fi} 
\def \Evir {\ifmmode E_{\rm vir} \else $E_{\rm vir}$ \fi} 
\def \lam {\ifmmode \lambda  \else $\lambda$ \fi} 
\def \lamp {\ifmmode \lambda^{\prime} \else $\lambda^{\prime}$  \fi} 
\def \Vmax {\ifmmode V_{\rm  max} \else  $V_{\rm max}$  \fi}

\title[Cores in warm dark matter haloes]{Cores in warm dark matter haloes: a {\it Catch 22} problem}

\author[A.V. Macci\`o \etal]
{Andrea V. Macci\`o$^{1}$\thanks{E-mail: maccio@mpia.de}, Sinziana Paduroiu$^2$, Donnino Anderhalden$^3$
\newauthor{Aurel Schneider$^3$, Ben Moore$^3$}\\
$^1$ Max-Planck-Insitute for Astronomy, K\"onigstuhl 17, D-69117 Heidelberg, Germany\\
$^2$ Geneva Observatory, University of Geneva, CH-1290 Sauverny, Switzerland\\
$^3$ Institute for Theoretical Physics, University of Z\"urich, CH-8057 Z\"urich, Switzerland
}

\date{\today}
\begin{document}
\maketitle

\begin{abstract}

The free streaming of warm dark matter particles dampens the 
fluctuation spectrum, flattens the mass function
of haloes and set a fine grained phase density limit for dark matter structures.
The phase space density limit is expected to
imprint a constant density core at the halo center on the contrary to what happens for
cold dark matter.
We explore these effects using high resolution simulations of structure 
formation in different warm dark matter scenarios.
We find that the size of the core we obtain in simulated haloes is in good
agreement with theoretical expectations based on Liouville's theorem. 
However, our simulations show that in order to create a significant core, ($r_c\sim 1$ kpc),
in a dwarf galaxy ($M\sim 10^{10} \Msun$), a thermal candidate with a mass as low as 0.1 keV is required.
This would fully prevent the formation of the dwarf galaxy in the first place.
For candidates satisfying large scale structure constrains ($m_{\nu}$ larger than $\approx 1-2$ keV) 
the expected size of the core is of the order of 10 (20) pc for a dark matter halo
with a mass of 10$^{10}$ (10$^8$) $\Msun$
We conclude that ``standard'' warm dark matter is not viable solution for explaining
the presence of cored density profiles in low mass galaxies.

\end{abstract}


\section{Introduction}

The formation of structure in the universe is driven by the mysterious dark matter component whose nature is still unknown. Over the last decades, 
the hierarchical cold dark matter model (CDM) has become the standard description for
the formation of cosmic structures. It is in excellent agreement with recent observations, such as measurements of the cosmic microwave background and large scale 
surveys \citep[][]{tegmark2006,komatsu2011}. However, there are a number of inconsistencies on sub-galactic scales that arise within the CDM scenario.
Firstly, the amount of substructure in Milky Way sized haloes is overpredicted by roughly one order of 
magnitude \citep[][]{klypin99,moore99}. Secondly, the central densities of CDM haloes in simulations show a cuspy 
behavior \citep[][]{moore94,flores_primack94,diemand05,Maccio07,springel08}, 
whereas the density profiles inferred from galaxy rotation curves point to a core like structure \citep[e.g.][]{deBlok01,deNaray2009,Oh2011}.
Furthermore, recent studies \citep{tikhonov09,zavala09,peebles2010} re-emphasized that also the 
population of dwarf galaxies within voids is in strong contradiction with CDM predictions.

One possible solution to these issues is that the dark matter particle is a thermal relic 
with a mass of order one keV. The most prominent representatives of such warm dark matter 
(WDM) candidates are the sterile neutrino and the gravitino
\citep[][]{aba06,boyarsky09a}, whose presence is also motivated by particle theory \citep[e.g.][]{dodelson_widrow94,buchm07,takayama2000}. 

Non-zero thermal velocities for WDM particles lead to a strong suppression of the linear matter power 
spectrum on galactic and sub-galactic scales 
\citep{bond_etal,pagels_primack82,dodelson_widrow94,hogan_dalcanton00,zentner03,abazajian06,viel05}, 
and erase all primordial density perturbations smaller than their free-streaming scale $\lambda_{fs}$.
Below this scale  no structure is expected to form, at least not in the usual bottom-up scenario. 
However, the effective suppression of halo formation already happens well above $\lambda_{fs}$ and is entirely described by 
the WDM particle mass \citep[see][and references therein]{smith_markovic11}. 

Recent observational constraints coming from X-ray background measurements and Ly-$\alpha$  forest analysis 
set the allowed mass interval roughly  between 2 and 50 keV
\citep[e.g.][]{viel05,seljak06,aba06,boyarsky09b,boyarsky09c}\footnote{In some of these analysis the warm dark matter particle is assumed to be a resonantly produced 
sterile neutrino \citep[][]{shifuller99}. We have converted these mass limits into limits for a fully thermalized particle, such as the gravitino, using the formula
provided by \citet[][]{viel05}.}
%
As a complementary study Macci\`o \& Fontanot (2010, see also Polisensky \& Ricotti 2011) compared the subhalo abundance of a 
Milky Way like object in different numerical warm dark matter realizations with observed satellite galaxies reported by 
the SDSS data and set a lower bound for a thermalized particle of $m_{\rm{WDM}} \gtrsim 2$ keV.

Another important characteristic of a WDM scenario is the possibility to {\it naturally} obtain cored
matter density profiles.
According to Liouville's theorem for collisionless systems, the fine grained phase space density of the 
cosmic fluid stays constant through cosmic history.
In WDM the dark matter fluid is described by a Fermi-Dirac distribution, whose absolute value is fixed at the time 
of decoupling when the fluid becomes collisionless. Structure formation then happens through a complex process of 
distortion and folding of the phase space sheet. Since it is not possible to measure this fine-grained phase space 
density in simulations, one usually defines a a coarse-grained or pseudo phase space density \citep[e.g.][]{tn01}
\begin{equation}
Q \equiv \frac{\rho}{\sigma^{3}},
\end{equation}
where $\rho$ is the mean density and $\sigma$ is the one-dimensional velocity dispersion within some small patch of the simulation\footnote{In the context of an non-singular isothermal sphere, the quantity Q is directly proportional to the maximum phase space density
and can be described, as in \citet{tg79} as giving the maximum coarse-grain phase space density. 
In a more general context, applicable to simulations, the velocity distribution of the particles is not Maxwellian and hence
$Q$ does not really trace the coarse-grained phase space density and hence we will refer to is as to {\it a pseudo phase space density}.}.
The quantity $Q$ corresponds to an average density of a small (but not microscopic) phase space volume and is not constant anymore. 
However, because of the way the phase space sheet is distorted, the value of $Q$ in most of the cases can only decrease during structure formation 
and will not exceed its  initial value set at decoupling (Dalcanton \& Hogan 2001, see however Boyarsky \etal 2009c for a thorough discussion
of the meaning of $Q$ and its evolution with time).

This upper limit for $Q$ also holds for the local pseudo phase space density within virialised  haloes at redshift zero and 
has a direct consequence on the density profile in real space.
Since the velocity dispersion does not grow in the inner part of 
a halo, the real space density profile must become constant with a core size depending on the specific WDM model \citep[][]{tg79}.

Due to this effect of core formation, the WDM scenario has been suggested as a solution to the long standing core-cusp problem of dwarf galaxies. In fact,
observational measurements favor cored dark matter profiles in 
low surface brightness galaxies within the local group \citep[][]{salucci2011, deNaray}. 
However, previous theoretical/analytical studies \citep[e.g.][]{devega10} argue that the cores produced by warm dark matter 
might be too small to explain the observations.
For example \citet{Bode01} argued that the principal effect of the thermal motion in the WDM scenario, is to give the particle angular momentum, 
producing a centrifugal barrier keeping	the particle away from	$r=0$; only for radii inside this barrier is the structure 
of the halo significantly altered with respect to a a pure CDM halo. Assuming a flat rotation curve for the halo and spherical collapse
they estimated that for warm particles with masses larger than 1 keV, thermal velocities are not able to
modify the structure of halos on scales of a kiloparsec or above.

More recently \citet{villa} have employed the spherical collapse model 
to study the formation of halos in warm dark matter cosmologies.
They found that the core sizes, for allowed WDM temperatures ($\sim 1$ keV), are typically very small, of the order to
$10^{-3}$ of the halo virial radius at the time of formation, and considerably smaller following formation.
They concluded that for realistic WDM models the core radii of halos observed at $z=0$ are generically expected
to be far smaller than the core sizes measured in local Low Surface Brightness galaxies.
One of the aim of our work is to test these previous analytical results 
using self-consistent cosmological N-body simulations of halo formation in a WDM universe.

Numerical N-body simulations have been used to better understand the properties of virialized objects in the Warm Dark Matter scenarios
\citep[e.g.][]{Bode01,knebe03,Wang,zavala09,tikhonov09,schneideretal11}. High resolution simulations of single 
objects have studied the suppression of the
galactic satellite formation due to free streaming \citep[e.g][]{colin2000,gotz,knebe08,maccio10b}, in order to reconcile the 
observed dwarf galaxy abundance with the prediction from Dark Matter based theories.
More recently \citet{colin08} used N-body simulations to study the effects of primordial (thermal) velocities on the inner structure of 
dark matter haloes, with particular attention on the formation of a possible central density core.
They used thermal velocities of the order of 0.1 and 0.3 km/s, without linking them to any particular WDM model, since the aim of their work was to
explore the general effect of relic velocities of the DM structure.
Unfortunately their combination of resolution and choice for relic velocities was not sufficient to directly test simulation results against
core radii predicted by phase-space constraints.

In this work we want to extend and improve on these previous studies.
We will use high resolution N-body simulations to explore the sizes of density cores 
in WDM and their dependence on the WDM candidate mass\footnote{In the present work we only considered a very simple WDM model;
it is worth commenting that there are more complex and physically  motivated models discussed in the literature (e.g. warm+cold dark matter,
\citet{boyarsky09d,maccio12b} or composite dark matter \citet{khlopov06,khlopov08})}.
We will explore several models for WDM ranging from 2 to 0.05 keV. 
We will consider  separately the effects of a warm dark matter candidate on the power spectrum and 
on the relic velocities, trying to disentangle the various consequences of these two different components.
Our higher numerical resolution will allow us to directly see the formation of a density core, with a size 
well above the numerical resolution for the warmer candidates.
We will then revise the theoretical arguments for the formation of cored profiles in WDM and perform
a direct comparison between the core sizes in our simulations and the ones predicted from phase space constraints.

The paper is structured as follows: In section 2 we discuss the setup of our simulations 
and the way we implement thermal velocities. Section 3 is dedicated to the presentation of our results 
in terms of the phase space limit and its influence on the density profile of dark matter haloes. 
A conclusion and a summary of our work is finally given in section 3. 

\section{Simulations}

Numerical simulations have been carried out using  {\sc pkdgrav}, a treecode
written by Joachim Stadel and Thomas Quinn \citep[][]{stadel01}. The initial
conditions are generated with the {\sc grafic2} package \citep[][]{bertschinger01}.
All simulations start at redshift $z_i=99$ in order to ensure a proper treatment
of the non linear growth of cosmic structures.

The cosmological parameters are set as follows: $\Omega_{\Lambda}$=0.727,
$\Omega_m$=0.273, $\Omega_b$=0.044, $h=0.7$ and $\sigma_8=0.8$, 
and are in good agreement with the recent WMAP mission 
results \citep[][]{komatsu2011}.

\begin{figure*}
\label{fig:50}
\begin{tabular}{ccc}
\psfig{file=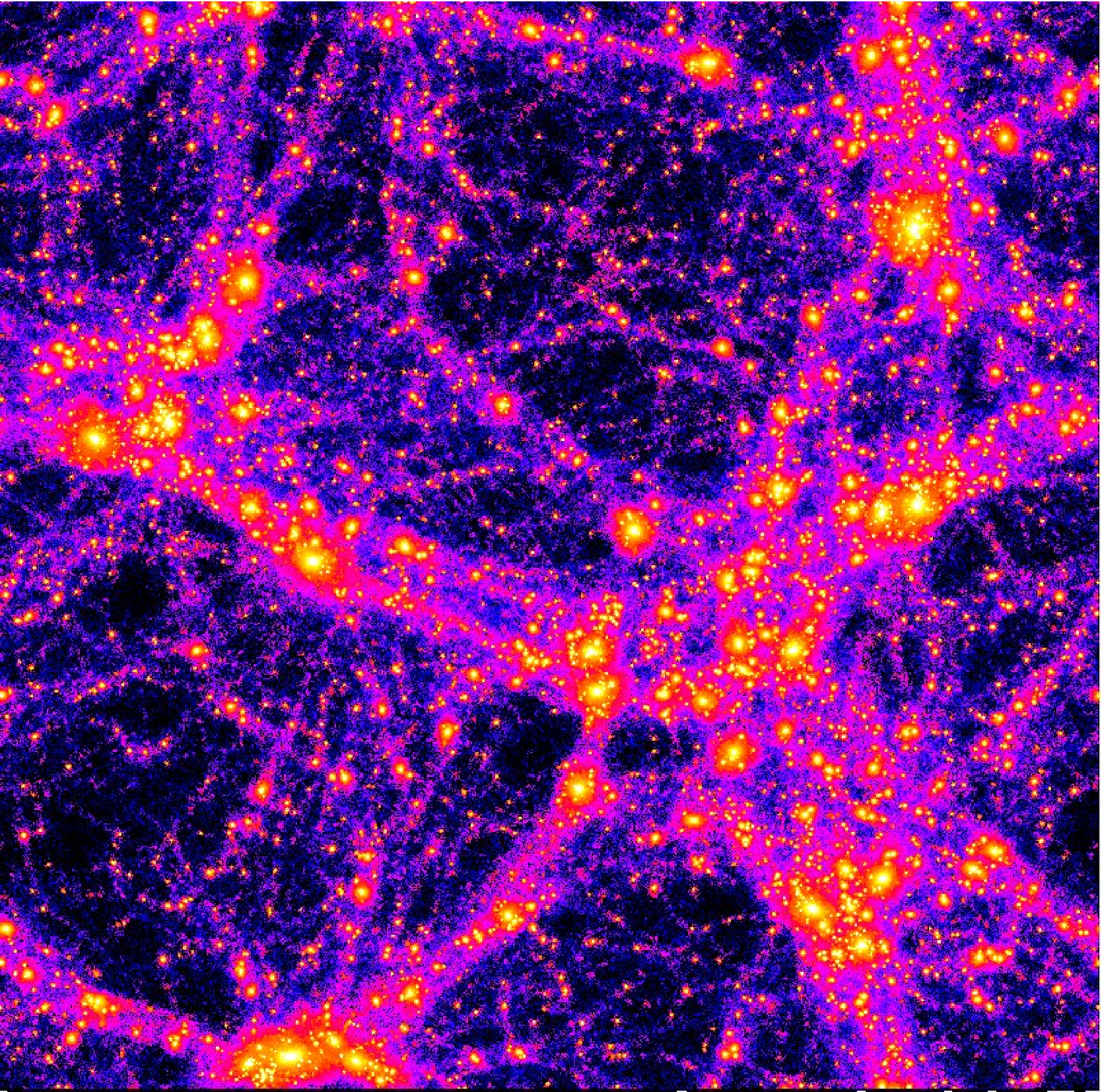,width=160pt} &
\psfig{file=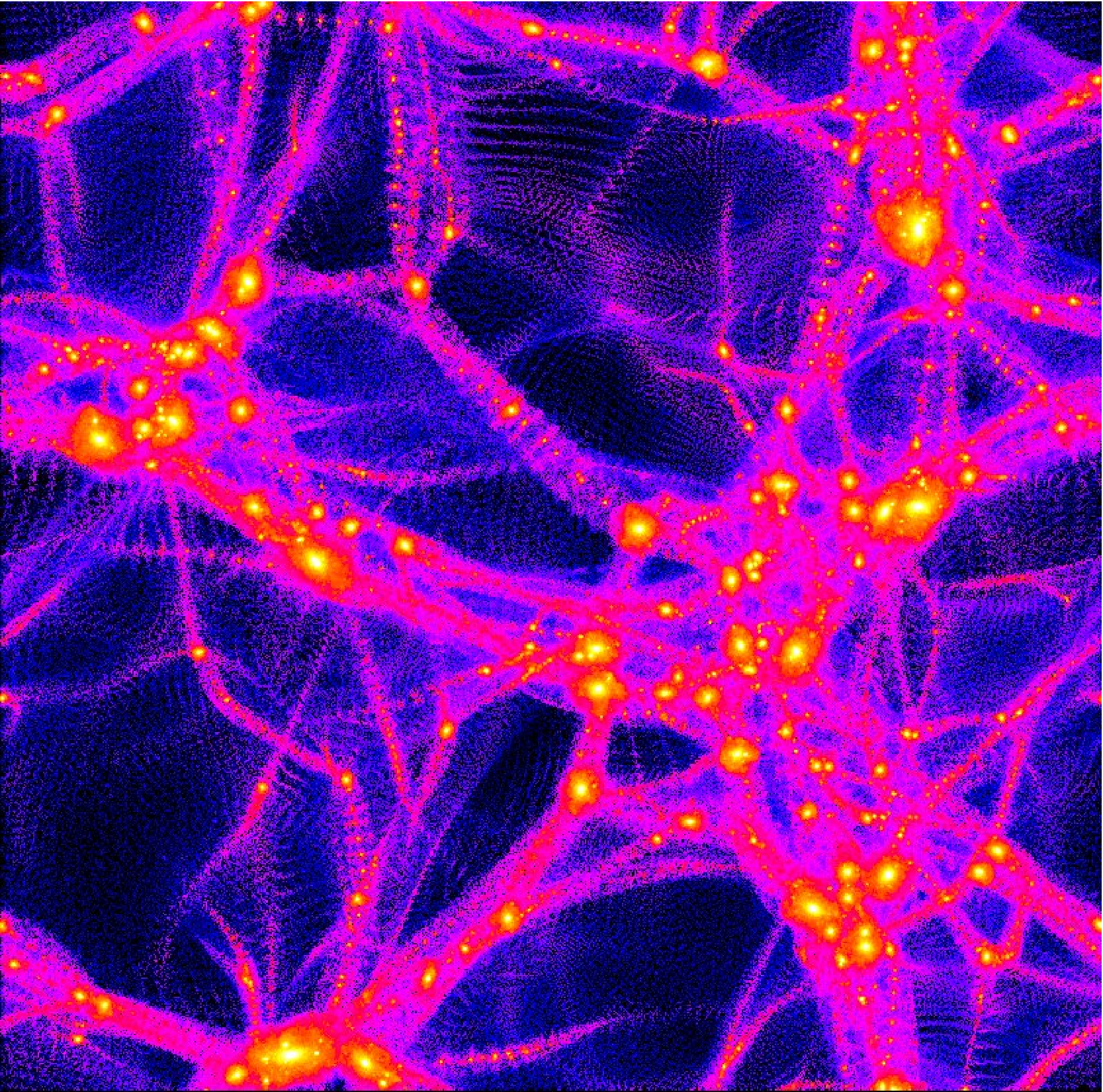,width=160pt} &
\psfig{file=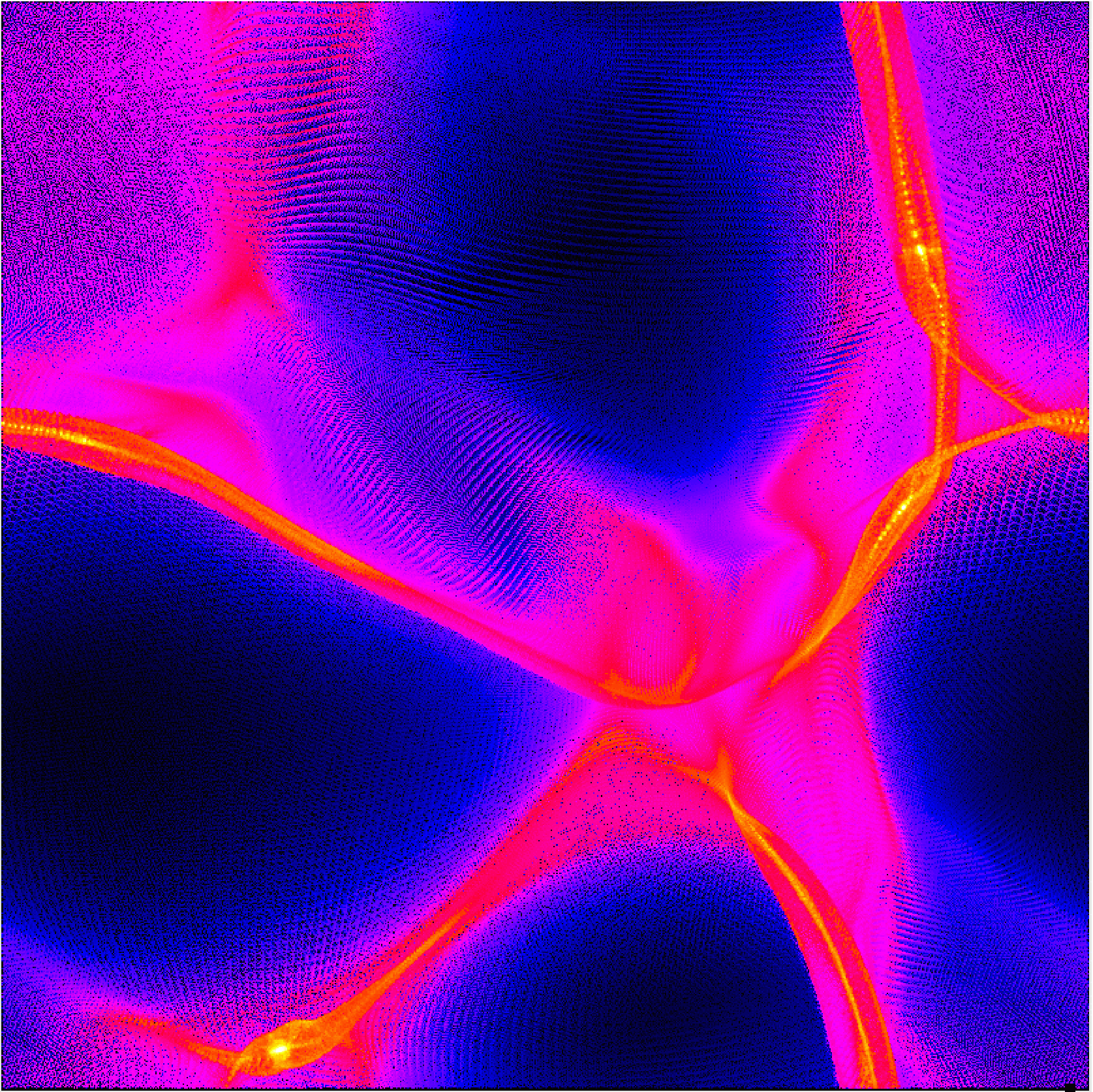,width=160pt} \\ 
\end{tabular}
\caption{Density map of the large scale (low resolution) simulations (L=40 Mpc) at redshift zero. 
From left to right: CDM, and two WDM with a cut in the power spectrum 
for a $m_{\nu}$ mass of 0.2 and 0.05 keV respectively. These last two simulations 
have not been used in this paper and are presented only for illustration purposes, see section
\ref{ssec:vel} for more information.} 
\end{figure*}

We start by running large scale simulations
of a cosmological cube of side 40 Mpc, using $2\times 256^3$ dark matter particles.
This was done for two different models, a standard LCDM and a Warm Dark Matter model
with a warm candidate of mass 2 keV produced in thermal equilibrium.

To compute the transfer function for WDM models we used the  
fitting formula suggested by \citet[][]{Bode01}:   
\begin{equation}   
T^2(k) = {P^{\rm{WDM}} \over P^{\rm{CDM}} } = [1+(\alpha k )^{2 \nu}]^{-10/\nu}   
\label{eq:Twdm}   
\end{equation}   
\noindent   
where $\alpha$, the scale of the break, is a function of the WDM parameters, while the index  
$\nu$ is fixed. \citet[][]{viel05} \citep[see also][]{hansen02}, using a Boltzmann code
simulation, found that $\nu=1.12$ is the best fit for $k<5 ~h~ \rm Mpc^{-1}$,  
and they obtained the following expression for $\alpha$:   
\begin{equation}   
\alpha = 0.049  \left ( {m_x \over  \rm{1 keV}}  \right )^{-1.11}   \left ( { \Omega_{\nu}  
\over 0.25 }\right )^{0.11}   \left ( { h \over 0.7} \right )^{1.22} \mpch.   
\label{eq:alpha}
\end{equation}   
We used the expression given in Eq. \eqref{eq:alpha} for the damping of the power-spectrum
for simplicity and generality. More accurate expressions for the damping of sterile neutrinos exist \citep[e.g.][]{abazajian06} 
and show that the damping depends on the detailed physics of the early universe
in a rather non-trivial way.
The initial conditions for the two simulations have been created using the same random phases, 
in order to facilitate the comparison between the different realizations.

We then select one candidate halo with a mass similar to our Galaxy ($M \sim 10^{12} \Msun$)
and re-simulated it at higher resolution.
These high resolution runs are 8$^3$ times more resolved in mass than the
initial ones: the
dark matter particle mass is $m_{p} = 1.38 \times 10^5 \Msun$, where 
each dark matter particle has a spline gravitation softening of 355 pc. 
This single halo has been re-simulated in several different models, all simulations are summarized 
in Table \ref{table:sim} and three of the simulations are shown in Figure 1.

\begin{table}
\begin{center}
\caption{Simulations parameters}
\label{table:sim}
\begin{tabular}{lccccc}
\hline
\hline
Label & $m_{\nu}$  & $m_{\nu, {\rm vel}}$ & $v_0(z=0)$ & $N_{vir}$ &  $M_{vir}$  \\ 
 & (keV) & (keV)  & (km/sec) &(10$^6$) &  ($10^{12} \Msun$)  \\ 
\hline
CDM & $\infty$ & -- & -- & 10.2  & 1.42 \\
\hline
WDM1 & 2.0 &  2.0   & 4.8 $\times 10^{-3}$  & 8.6     & 1.22 \\
WDM2 & 2.0  & 0.5   & 3.1 $\times 10^{-2}$  & 8.4      & 1.20   \\
WDM3 & 2.0  & 0.2   & 0.1     & 8.5     & 1.21   \\
WDM4 & 2.0  & 0.1   & 0.26     & 6.7     & 0.93  \\
WDM5 & 2.0  & 0.05  & 0.66     & 4.9   & 0.71  \\
\hline
\hline
\hline
\end{tabular}
\end{center}
\end{table}

\subsection{Streaming velocities}
\label{ssec:vel}

Particles that decouple whilst being relativistic are expected to retain a thermal velocity component.
This velocity can be computed as a function of the WDM candidate mass ($m_{\nu}$) according to the following
expression \citep[][]{Bode01}:

\begin{equation}
\label{eq:velocities}
\frac{v_0(z)}{1+z}=.012\left(\frac{\Omega_{\nu}}{0.3}\right)^{\frac{1}{3}}\left(\frac{h}{0.65}\right)^{\frac{2}{3}}\left(\frac{1.5}{g_{X}}\right)^{\frac{1}{3}}\left(\frac{\text{keV}}{m_{\nu}}\right)^{\frac{4}{3}}\kms
\end{equation}
where $z$ is the redshift. The distribution function is the given by the Fermi-Dirac expression 
until the gravitational clustering begins \citep{Bode01}.

This formalism is correct for the ``real'' dark matter elementary  particles (e.g. a sterile neutrino).
In the N-body approach we use macro particles (with masses of the order of $10^5 \Msun$) 
to describe the density field. These macro particles effectively model a very large number of micro particles.
Given that the velocities described in Eq. \eqref{eq:velocities} have a random direction
the total velocity of the macro (N-body) particles should effectively be zero.
Hence, it is not fully correct to directly use Eq.  \eqref{eq:velocities} to assign 
``thermal'' velocities to simulation particles.

On the other hand, the net effect of the thermal velocities is to create a finite upper limit
in the phase-space distribution (PSD) due to their initial velocity dispersion ($\sigma$).
What we are interested in is to recreate the same PSD limit in our simulation, and then 
study its effects on the dark matter halo density distribution.
In order to achieve this goal we proceed in the following way.
From Eq. \eqref{eq:velocities} we compute the rms velocity: $\sigma(z)=3.571 v_0(z)$, we
then create a Gaussian distribution centered on zero and with the same rms $\sigma$.
Finally we randomly generate particle velocities from this distribution and assign them 
to our macro particles.
It is worth mentioning that the final results are almost independent on the 
assumed distribution for the velocities (Fermi-Dirac, Maxwellian etc.), while they strongly depend
on the strength of the velocity field (i.e. $v_0$).

In principle  adding random velocities introduces spurious momentum fluctuations
into the initial conditions. For very light particles ($m_{\nu}\sim 1$ eV) this effect
could be important and it could be balanced by 
introducing particles with opposite momenta (e.g. Gardini, Bonometto \& Murante 1999).
On the other hand, for the choices of WDM candidate masses in our paper, thermal velocities are quite modest
($\lta$ 0.5 km/sec) and lower than the Zeldovich ones. So no artificial effects are expected.

As detailed in section \ref{ssec:theory}, there is a direct connection between $m_{\nu}$
and the expected size of the dark matter distribution core. This core is only due to the presence of 
thermal velocities and not, in first approximation, to the cut in the power spectrum described by Eq. \eqref{eq:Twdm}. 
Cutting the power spectrum changes the merger history of the dark matter halo but does not affect the density profile significantly \citep[][]{moore99}.
This implies that in order to study the effect of different values of $m_{\nu}$ (and hence $v_0$) it is sufficient
to {\it ``play''} with Eq. \eqref{eq:velocities} leaving all other simulation parameters unaltered.
Following this approach we have generated several simulations using the same cut in the power spectrum
($m_{\nu}$) but different initial thermal velocities ($m_{\nu, {\rm vel}}$), as detailed in Table 1.

\section{Results}

\begin{figure}   
\psfig{file=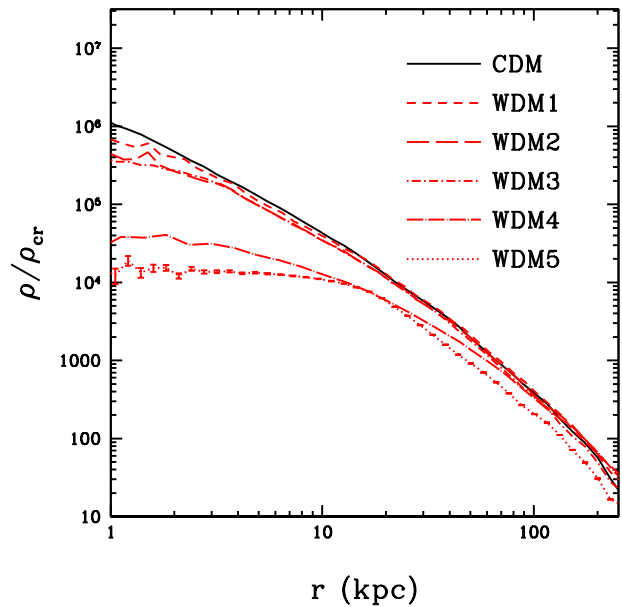,width=0.5\textwidth}
\caption{The spherically averaged density profiles for CDM, WDM1-5 haloes.}
\label{fig:prof}
\end{figure}

Density profiles for the cold dark matter run and the four warm dark matter realizations
(WDM1-WDM5) are shown in Fig. \ref{fig:prof}.
The profiles show a monotonic decrease of the central density as a function of the 
the temperature of the dark matter candidate. Cold candidates show the usual cuspy 
behavior \citep[e.g.][]{dc91}, while warmer candidates present a lower central density that becomes a clear
core for $m_{\nu, {\rm vel}} = 0.05$ keV, with a size of several kpc.

\begin{figure}   
\psfig{file=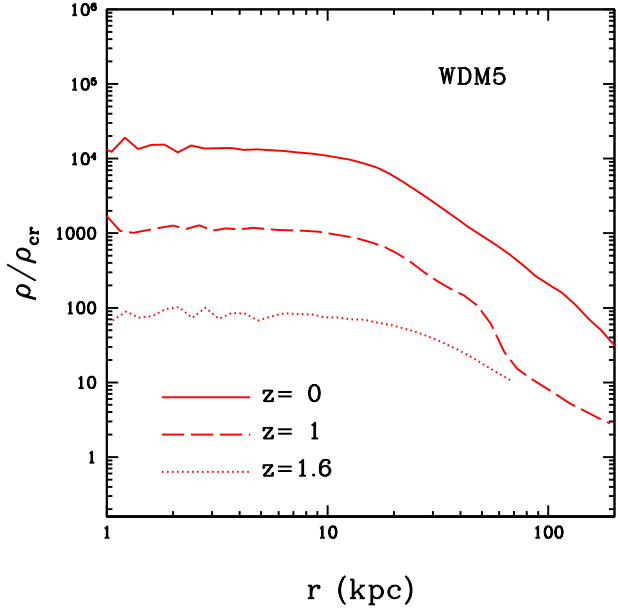,width=0.5\textwidth}
\caption{Time evolution of the density profiles for the WDM5 halo.}
\label{fig:profz}
\end{figure}

Fig. \ref{fig:profz} shows the time evolution of the density profile in the WDM5 simulation.
The profile is already cored at high redshift $z=1.6$, and the size of the core does not evolve substantially 
until $z=0$. The profile only changes at large radii ($r> 50$) kpc, as a consequence 
of the assembly of the external part of the halo. This smooth mass accretion is also a consequence 
of the quiet merging history of the halo, that does not undergo any merger with a mass ratio
larger than 10 after z=2.  The assembly of the external part of the halo is consistent 
with a typical CDM halo in the outer regions.

\begin{figure}   
\psfig{file=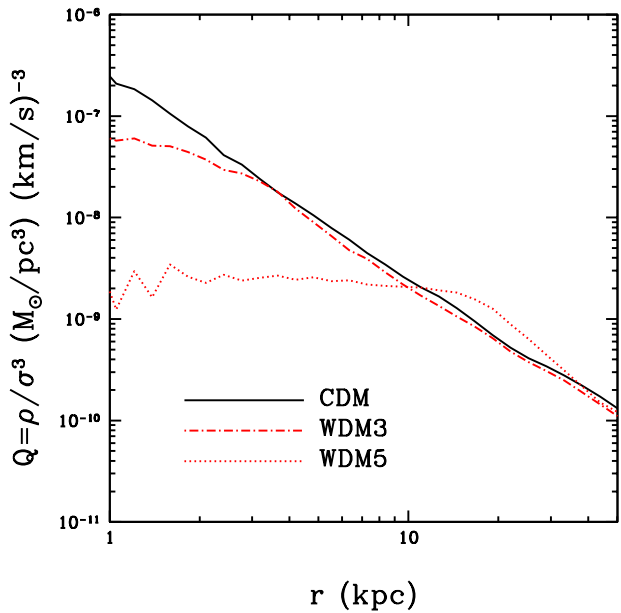,width=0.5\textwidth}
\caption{Phase space density profile for the CDM, WDM3 and WDM5 models at $z=0$.}
\label{fig:qmax}
\end{figure}

As already mentioned the theoretical explanation for the formation of a core is related to 
the presence of a maximum in the phase space density distribution.
This maximum is clearly visible in Fig. \ref{fig:qmax}, where we plot the pseudo phase space density  
$Q \equiv \rho/\sigma^3$, for three different models, namely CDM, WDM3 and WDM5.
For this latter model the $Q$ shows a large core that extends about 10 kpc.
The WDM3 model also shows a strong flattening of the $Q$ profile, consistent with a core distribution.
On the other hand the CDM pseudo phase-space distribution is well fitted by a single power law profile
on the whole range, in agreement with previous results \citep[][]{tn01,schmidt08}.

\begin{figure}   
\psfig{file=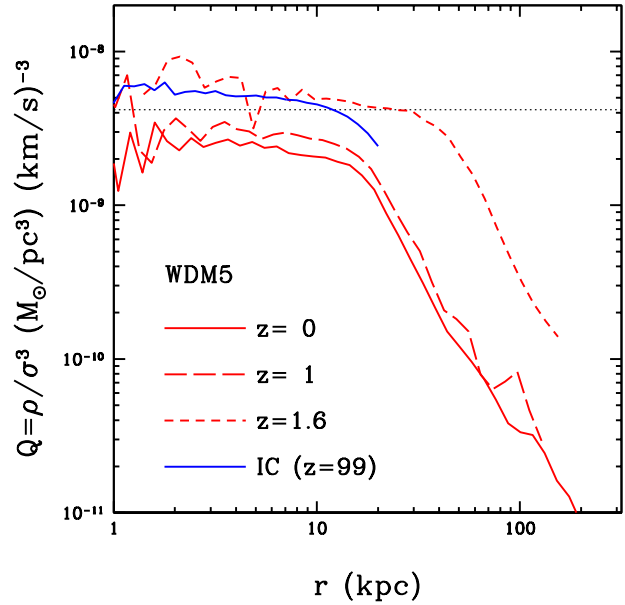,width=0.5\textwidth}
\caption{Time evolution of the pseudo PSD radial profile for the WDM5 model. The black dotted line represents the theoretical prediction 
for the maximum value of $Q$ according to equation \ref{eq:qmax}.}
\label{fig:Qmaxz}
\end{figure}

Figure \ref{fig:Qmaxz} shows the time evolution of the pseudo PSD for our warmest candidate 
(i.e. thermal velocities for a 0.05 keV mass particle).
The solid (blue) line shows the $Q$ radial profile in the initial conditions ($z=99$).
This value has been calculated using only high resolution particles that end up within 
1.5 times the virial radius of the halo at $z=0$.
The other (red) lines represent the pseudo PSD profile
at different redshifts (from 1.6 to 0) and have been computed using all particles within 
the virial radius of the halo. All quantities in the plot are in physical units.
The phase space distribution shows very weak evolution with almost no evolution
at all from $z=99$ to $z=1.6$. 
In the same plot we also show the theoretical maximum phase-space density
achievable by this model (see Eq. \eqref{eq:qmax} for a rigorous definition of $Q_{max}$).

The dotted (black) lines show predictions for 
$Q_{max}$ for the {\it local} value of the matter density, which we measured directly from the simulation 
initial conditions using DM particles in the high resolution region within a volume of $\approx 1$ Mpc$^3$.
The local density value turned out to be $\langle \rho \rangle_{local}=0.31 \times \rho_{cr}$
\footnote{This local value is slightly higher than the global one since it is computed 
around an object that will collapse and be fully virialized at $z=0$}.
The theoretical prediction is in quite good agreement with the simulation results.

In order to quantify the flatness (and the core size) of WDM profiles we have
fitted all our density profiles with the following parametric description, originally presented in \citet[][]{stadel09}:
\begin{equation}
\rho(r) = \rho_0 \exp (-\lambda [ \ln (1+r/R_{\lambda})]^2).
\label{eq:sm}
\end{equation}

In this parameterization the density profile is linear down to a scale $R_{\lambda}$ beyond which it approaches 
the central maximum density $\rho_0$ as $r \rightarrow 0$. We also note that if one makes a plot of 
$\rm {d ln} \rho/ {\rm d ln (1 + r/R_{\lambda})}$ versus $ \rm{ln} (1 + r/R_{\lambda})$ 
then this profile forms an exact straight line with slope $2\lambda$. 

\begin{figure}   
\psfig{file=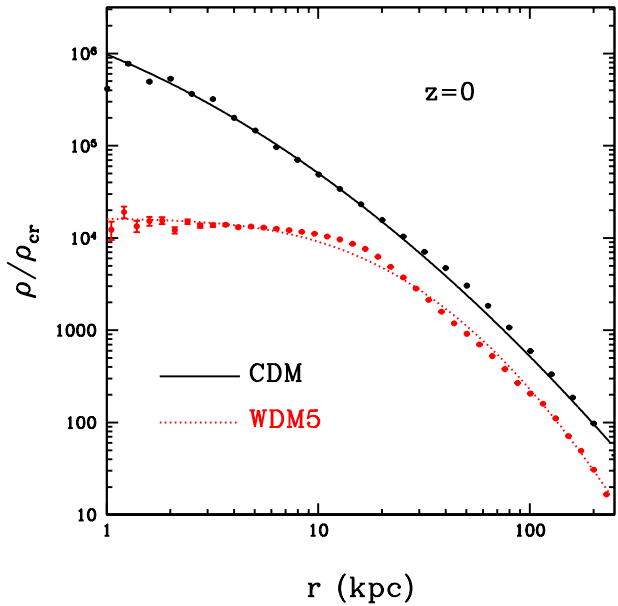,width=0.5\textwidth}
\caption{Density profiles for CDM and WDM5 and their fit using Eq. \eqref{eq:sm}.}.
\label{fig:sm}
\end{figure}

This fitting function is extremely flexible and makes possible to reproduce at the same time both cuspy 
profiles like the one predicted by the CDM theory, as well as, highly cored profiles, like in the WDM5 case 
(as shown in Fig. \ref{fig:sm}).
The values of the parameter are obtained via a  $\chi^2$ minimization  procedure  
using the  Levenberg \&  Marquart method.
From now on we will use the value of the fitting parameter $R_{\lambda}$ as the fiducial value
of the central density core  in simulated profiles ($r_{\rm core,s}$, hereafter). The $r_{\rm core,s}$  values for all our haloes are reported in the first
column of Table 2.

\subsection{Comparison with theoretical predictions}
\label{ssec:theory}

In \citet[][TG79 hereafter]{tg79} limits on the mass of a neutrino are derived from
the maximum phase space density of a homogeneous
neutrino background, with the further assumptions that
neutrinos form bound structures and that their central regions can
be well-approximated by an isothermal sphere. 

Assuming a Maxwellian velocity distribution they obtained the maximum phase space density:
\begin{equation}
Q_{max} \equiv { \rho \over \sigma^3} \propto m_{\nu}^{4}
\end{equation}
\noindent
where $m_{\nu}$ is the mass of the (warm) dark matter candidate.
This limit has been then used in several follow up papers to estimate the size of density cores
in warm dark matter haloes \citep[e.g.][]{dalcanton_hogan01,strigari}.

Following TG79 we derive the theoretical expectation for the maximum pseudo-phase space density 
and the size of the DM core for our WDM models adopting a slightly different approach.
We can start from the definition of $Q$ assuming to compute the density in some local volume $L$:
\begin{equation}
Q_{max} \equiv { \rho_L \over \sigma^3} = { {{\rho_L} \over  {\rho_{cr}}} \times \rho_{cr} \over \sigma^3}
\label{eq:qmax}
\end{equation}
where $\rho_{cr}= 2.775 \times 10^{11} h^2 \Msun \rm {Mpc^{-3}}$ is the critical density of the Universe and
$\rho_L/\rho_{cr}$ is the local density in our volume $L$, expressed in units of the critical density.

The denominator of Eq.\ref{eq:qmax} could be expressed as a function of the mass of the WDM candidate using
Eq.\ref{eq:velocities} and the fact that for a Fermi-Dirac distribution the rms velocity is $\sigma=3.571*v_0$.
Combining Eq.\ref{eq:velocities} with Eq.\ref{eq:qmax} we get the following expression for $Q_{max}$:

\begin{equation}
Q_{max} = 1.64 \times 10^{-3} 
\left( {{\rho_L} \over  {\rho_{cr}}} \right)
\left( {m_{\nu} \over  {\rm keV} } \right)^{4} 
{ {\Msun{\rm pc}^{-3}} \over \left(\kms\right)^3 }.
\label{eq:qmax1}
\end{equation}
Where the numerical factor in front of the expression takes into account our choices for $\Omega_m$ and $h$.
This expression is formally equivalent to the one derived by TG79.

Finally the maximum phase space density can be converted in a 'core' size following \citet[][]{hogan_dalcanton00}:
\begin{equation}
r_{\rm core,t}^{2}=\frac{\sqrt{3}}{4\pi  GQ_{max}}\frac{1}{<{\sigma_{halo}^{2}>}^{1/2}}.
\label{eq:core}
\end{equation}
where $\sigma_{halo}$ is the velocity dispersion (i.e. the mass) of the simulated dark matter halo.
Values of $r_{\rm core,t}$ for our simulated haloes are reported in the last column of Table \ref{table:core}.

In the following we will compare this theoretical value of the core ($r_{\rm core,t}$) with two different core sizes than 
can be estimated directly from the simulations. The first one is given by the $R_{\lambda}$ parameter
obtained by fitting the numerical density profile, as shown in Fig. \ref{fig:sm} and we will refer to this value
as $r_{\rm core,s}$. The second one is obtained by computing $Q_{max}$ from the simulated density profile (as shown in Fig. 
\ref{fig:qmax}) and then inserting this value in Eq. \eqref{eq:core}, we named this second parameter $r_{\rm core,Q}$.

Results for the three definitions of the core size for all our simulations are summarized in Table \ref{table:core}.
Overall the three different estimators for the core size are in fairly good agreement. $r_{c,Qmax}$ gives on 
average a larger value for the core, for the WDM4 and WDM5 runs, while for the WDM3 simulation is only 
able to give an upper value, since there is not a clear indication of convergence towards a maximum
value in the $Q_{max}$ profile, as shown in Fig. \ref{fig:qmax}.

\begin{table}
\begin{center} 
\caption{Size of density cores using different methods. See text for a more detailed explanation}
\label{table:core}
\begin{tabular}{lccc}
\hline
\hline
Label & $r_{\rm core, s}$  & $r_{\rm core, Q}$ & $r_{\rm core, t}$  \\ 
 & (kpc) & (kpc)  & (kpc)  \\ 
\hline
CDM &  $<0.4$ &  $<0.4$ & $\infty$ \\
\hline
WDM1 & $<0.4$ &  $<0.4$ & 0.005 \\
WDM2 & $<0.4$ &  $<0.4$ & 0.075 \\
WDM3 & 0.42   & $<1.1$ & 0.48 \\
WDM4 & 1.63   & 1.80   & 1.91  \\
WDM5 & 4.56   & 4.85   & 6.98 \\
\hline
\hline
\hline
\end{tabular}
\end{center}
\end{table}

\begin{figure}   
\psfig{file=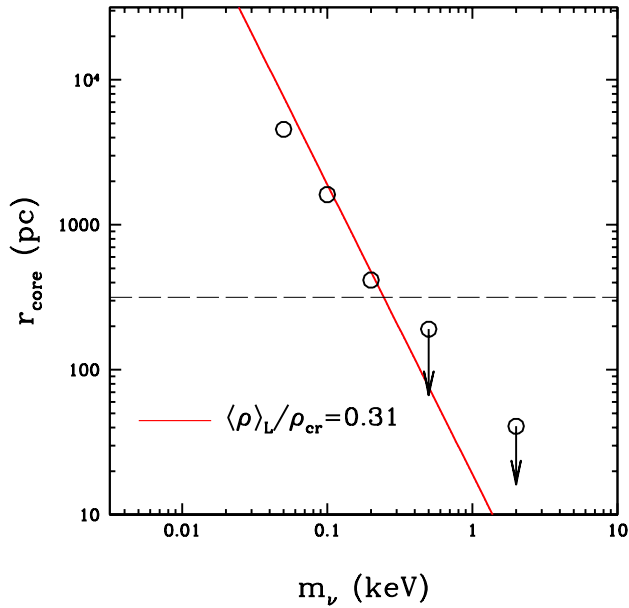,width=0.5\textwidth}
\caption{Comparison between core size in simulations (open symbols) and the theoretical expectation for 
a $M=10^{12} \Msun$ halo (solid line). The dashed line is the gravitational softening of our simulations. All points below
this line should be considered as upper limits on the core size.}
\label{fig:core}
\end{figure}

Fig. \ref{fig:core} shows the comparison of the core found directly in simulations ($r_{\rm core,s}$, black symbols) 
with the core predicted by the above simple theoretical argument ($r_{\rm core,t}$).
The solid line is obtained from Eqs. \eqref {eq:core} and \eqref{eq:qmax1}, 
where, as discussed before, we used $\rho_L/\rho_{cr}=0.31$ as value for the local density.

Overall numerical results for WDM3, WDM4 and WDM5 are in very good agreement with the theoretical expectations
from Eqs. \eqref {eq:core} and \eqref{eq:qmax1}.
The WDM1 and WDM2 simulations only put upper limits on the size of the core, since the values of 
$R_{\lambda}$ we obtain from fitting the density profile fall below the simulation softening (the dashed 
black line in the figure). 

\begin{figure}   
\psfig{file=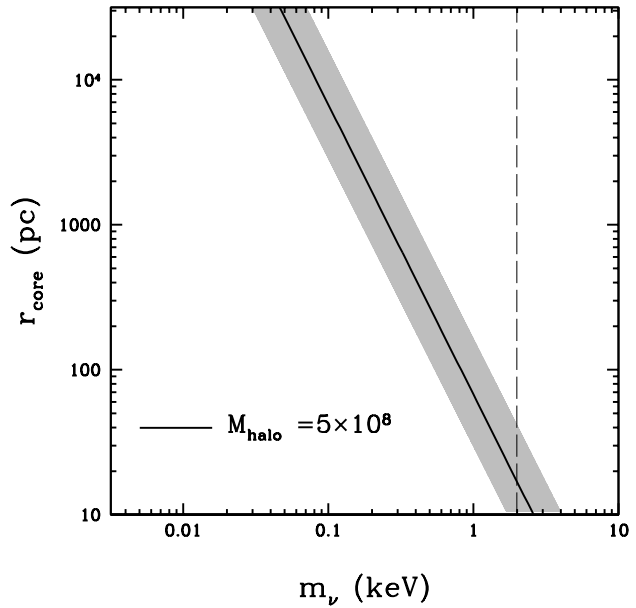,width=0.5\textwidth}
\caption{Expected core size for the typical dark matter mass of Milky Way satellites as a function of the WDM mass $m_{\nu}$.
The shaded area takes into account possible different values of the local density parameter  $0.15<\Omega_m<0.6$.
The vertical dashed line shows the current limits on the WDM mass from large scale structure observations.}
\label{fig:core2}
\end{figure}

Using our determination of the core size as a function of the warm dark matter mass we compute
the expected value of $r_{\rm core}$ for the typical halo mass ($5 \times 10^8 \Msun$, see \citet{maccio10a}) 
of dwarf galaxies orbiting the Milky-Way.
Results are shown in Fig. \ref{fig:core2}: the grey shaded area takes into account possible different values of the local
matter density in the range $\rho / \rho_{cr}= 0.15- 0.6$.

From the figure it is clear that a core of $\approx 1$ kpc would require a wdm mass of 
the order of 0.1 keV, well  below current observational limits from large scales.

If we assume a warm dark matter particle mass of $m_{\nu} \sim 2$ keV (represented by the dashed vertical line), in agreement
with several astrophysical constraints \citep[e.g.][]{viel08}, the maximum core size we can expect ranges
from 10 pc for a massive, MW-like halo (see also figure \ref{fig:core}), to 10-40 pc for a dwarf galaxy like halo.
Finally, in predicting the core size for satellite galaxies in the MW halo, it must be taken into account that
due to stripping and tidal forces satellites can lose significant mass after accreting into larger haloes \citep[e.g.][]{penarrubia08,maccio10a}. 
This implies that the halo mass we may infer today for those galaxies is only a lower limit on the mass they had 
before accretion, which is the one to be used (as $\sigma^2_{halo}$) in Eq. \ref{eq:core}.

\section{Conclusions}

We have used high resolution N-body simulations to examine the effects of free streaming velocities on halo internal structure in warm dark matter models.
We find:
\begin{itemize}

\item The finite initial fine grained Phase Space Density (PSD) is also a maximum of the pseudo PSD, 
resulting in PSD profiles of WDM haloes that are similar to CDM haloes in the outer regions, 
however they flatten towards a constant value in the inner regions. This is in agreement
with previous studies based on simulations \citep{colin08} and theoretical arguments
\citet{villa}.

\item The finite PSD limit results in a constant density core with characteristic size that is 
in agreement with theoretical expectations i.e. following Tremaine \& Gunn 1979, especially 
if value of the local matter density is taken into account.

\item The core size we expect for thermal candidates allowed by independent constraints on large scales
(Lyman-$\alpha$ and lensing, $m_{\nu} \approx 1-2$ keV), is of the order of 10-50 pc. This 
is not sufficient to explain the observed cores in dwarf galaxies that are around 
kpc scale \citep[][]{walker11,amorisco11,jardel12}.

\item Our results show that a core around kpc scale in dwarf galaxies, would require a thermal candidate with 
a mass below 0.1 keV,  ruled out by all large scale structure constraints \citep[][]{seljak06,miranda07,viel08}.
Moreover with such a warm candidate, the exponential cut-off of the Power Spectrum would make impossible 
to obtain these dwarf galaxies in the first place \citep[e.g.][]{maccio10b}. 

\item All together these results lead to a nice ``Catch 22'' problem for warm dark matter:
{\it If you want a large core you won't get the galaxy, if you get the galaxy it won't have a large core}.

\end{itemize}

We conclude that the solution of the cusp/core problem in local group galaxies 
cannot completely reside in simple models (thermal candidates) of warm dark matter. 
If cores are required then it seems that baryonic feedback \citep[e.g.][]{Romano08,governato10,maccio12a} is still the most likely 
way to alter the density profile of dark matter and hence 
reconcile observations with cold/warm dark matter predictions.

\section*{Acknowledgments}
We acknowledge stimulating discussions with George Lake, J\"urg Diemand and Justin Read.
We would like to thank A. Boyarsky, O. Ruchayskiy and S. Sergio Palomares-Ruiz for their help
with the theory part of this work.
Finally we thank the referee of our paper, Adrian Jenkins, for several comments that
improved the presentation and the clarity of our work.
Numerical simulations were performed on the {\sc theo} and PanStarrs2 clusters of 
the Max-Planck-Institut f\"ur Astronomie at the Rechenzentrum in Garching and on the
zBox3. S. P. would like to thank Joachim Stadel and Doug Potter for their crucial contribution to this project.
AVM acknowledges funding by Sonderforschungsbereich SFB 881 ``The Milky Way System'' 
(subproject A1) of the German Research Foundation (DFG).


\end{document}